\documentclass[reprint,showpacs,showkeys,noeprint]{revtex4-1} 
\usepackage[utf8]{inputenc}
\usepackage{cmap} 
\usepackage[T1]{fontenc}
\usepackage{microtype}
\usepackage{amsmath,amssymb,mathtools}
\usepackage{txfonts}

\usepackage{mathrsfs}

\usepackage{graphicx,grffile,textcomp}
\graphicspath{.}

\usepackage{textcomp}
\usepackage{booktabs,tabularx,dcolumn}
\newcolumntype{d}[1]{D{.}{.}{#1}}

\usepackage{url,hyperref}
\usepackage[usenames,dvipsnames]{xcolor}
\hypersetup{colorlinks=true, linkcolor=BrickRed, urlcolor=blue!50!black, citecolor=blue!50!black}

\begin{document}
\title{Aging Phenomena during Phase Separation in Fluids: Decay of 
autocorrelation for vapor-liquid transitions}
\author{Sutapa Roy$^{1,2}$, Arabinda Bera$^{3,4}$, Suman Majumder$^{3,5}$ and Subir K. Das$^{3,4,*}$}

\affiliation{
$^1$ Max-Planck-Institut für Intelligente Systeme,
   Heisenbergstr.\ 3,
   70569 Stuttgart,
   Germany\\
$^2$   Institut für Theoretische Physik IV,
   Universität Stuttgart,
   Pfaffenwaldring 57,
   70569 Stuttgart,
   Germany\\
$^3$ Theoretical Sciences Unit, Jawaharlal Nehru Centre for Advanced Scientific
Research, Jakkur P.O, Bangalore 560064, India\\
$^4$ School of Advanced Materials, Jawaharlal Nehru Centre for Advanced Scientific 
Research, Jakkur P.O, Bangalore 560064, India\\
$^5$ Institut für Theoretische Physik, Universität Leipzig, 
Postfach 100920, D-04009, Leipzig, Germany\\}
\date{\today}

\begin{abstract}
We performed molecular dynamics simulations to study relaxation phenomena during vapor-liquid 
transitions in a single component Lennard-Jones system. Results from two different overall 
densities are presented; one in the neighborhood of the vapor branch of the coexistence curve 
and the other being close to the critical density. The nonequilibrium morphologies, growth 
mechanisms and growth laws in the two cases are vastly different. In the low density case 
growth occurs via diffusive coalescence of droplets in a disconnected morphology. On the other hand, 
the elongated structure in the higher density case grows via advective transport of particles 
inside the tube-like liquid domains. The objective in this work has been to identify how the 
decay of the order-parameter autocorrelation, an important quantity to understand aging dynamics, 
differs in the two cases. In the case of the disconnected morphology, we observe a very robust 
power-law decay, as a function of the ratio of the characteristic lengths at the observation time 
and at the age of the system, whereas the results for the percolating structure appear rather complex. 
To quantify the decay in the latter case, unlike standard method followed in a previous study, here 
we have performed a finite-size scaling analysis. Outcome of this analysis shows the presence of 
a strong preasymptotic correction, while revealing that in this case also, albeit in the asymptotic 
limit, the decay follows a power-law. Even though the corresponding exponents in the two cases 
differ drastically, this study, combined with a few recent ones, suggests that power-law behavior of this 
correlation function is rather universal in coarsening dynamics. 

\end{abstract}
\maketitle

\section{Introduction}
Following a quench inside the coexisting curve a homogeneous system coarsens via the formation of 
domains of particle-rich and particle-poor regions \cite{bray,wadhawan,onuki,binderbook1001,jones,lifshitz,furukawa1,furukawa2,binder2,
binder3,tanaka1,tanaka2,das2011,das2012,roy2012,roy2013,roy2013jcp,jung2015,nielaba,watanabe,jung2016,
roy2018,tanaka2015,janke2017,azizi,bouttes,tanaka2000,laradji,lip,hugo,pinaki,yeungn,chris,xia,ana1,hsu,
vul,raz,sten,vin,jae1,jae2}. This coarsening is a self-similar phenomenon 
\cite{bray,wadhawan,onuki,binderbook1001,jones} with the average size, $\ell$, of the domains 
typically growing algebraically with time ($t$) \cite{bray,wadhawan,onuki,binderbook1001,jones}: 
\begin {eqnarray}\label{l_v_t}
\ell(t) \sim t^\alpha.
\end{eqnarray}
Universality of the exponent $\alpha$ depends, among other parameters, upon \cite{bray} the 
conservation and number of components of the appropriate order parameter, relevance of hydrodynamics, etc. 
In the case of ordering dynamics in a ferromagnet \cite{bray} or kinetics of phase separation in 
multi-component mixtures \cite{bray,wadhawan,binderbook1001,lifshitz}, with underlying state of 
the system being solid, space dimension ($d$) and initial composition play only minor role. 
However, these facts can drastically alter the value of $\alpha$ in fluids, where hydrodynamics is 
important \cite{bray,wadhawan,onuki}. Despite such weak universality, e.g., compared to equilibrium 
critical phenomena \cite{fisher1967}, significant understanding has been obtained with respect to 
the values of $\alpha$ in differing situations, via experiments, analytical theories as well as 
computer simulations \cite{bray,wadhawan,onuki,binderbook1001,jones,lifshitz,furukawa1,furukawa2,binder2,
binder3,tanaka1,tanaka2,das2011,das2012,roy2012,roy2013,roy2013jcp,jung2015,nielaba,watanabe,jung2016,
roy2018,tanaka2015,janke2017,azizi,bouttes,tanaka2000,laradji,lip,hugo,pinaki,yeungn,chris,xia,ana1,hsu,
vul,raz,sten,vin,jae1,jae2}. 

Another aspect that is important for the deeper understanding of such nonequilibrium processes is 
the aging property \cite{wadhawan,fisher1988,liu1991,majumdar1995,yeung,corberi2006,henkel,janke2007,das2014,
das2015,ahmad2012,das2013}. A crucial quantity in this context is the order-parameter autocorrelation 
function, $C_\text{ag}$. This quantity is a function of the observation time $t$ and the waiting time $t_w$, 
the latter being also often referred to as the age of the system, and is defined as \cite{wadhawan,fisher1988}
\begin {eqnarray}\label{aging1}
C_\text{ag}(t,t_w)= \langle \psi(\vec r,t)\psi(\vec r,t_w)\rangle-
\langle \psi(\vec r,t)\rangle \langle \psi(\vec r,t_w)\rangle,
\end{eqnarray}
the angular brackets representing statistical averaging. Here $\psi(\vec r,t)$ is an appropriate 
space ($\vec r$) and time dependent order parameter, which, for a vapor-liquid transition, can be 
defined as the local density fluctuation around a mean value. Decay of $C_\text{ag}$ probes the 
relaxation starting from different ages \cite{fisher1988}. 

Due to the violation of time translation invariance during growth processes, $C_\text{ag}$ is not 
expected to exhibit scaling as a function of $t-t_w$. Fisher and Huse (FH) \cite{fisher1988}, from 
the studies of spin-glass systems, proposed a scaling form of this quantity as 
\begin {eqnarray}\label{aging2}
C_\text{ag}(t,t_w) \sim \left ( \frac{\ell}{\ell_w}\right )^{-\lambda}.
\end{eqnarray}
Here $\ell$ and $\ell_w$ are the average domain lengths at times $t$ and $t_w$, 
respectively. FH also provided bounds on the exponent $\lambda$:
\begin {eqnarray}\label{aging3}
\frac{d}{2} \le \lambda \le d.
\end{eqnarray}

The scaling of $C_{\rm ag}$ as a function of $\ell/\ell_w$ has been observed to be valid in a 
number of phase ordering systems. An interest in the literature has been to learn on the 
universality of the scaling function. Some recent studies suggest \cite{das2014,das2015} that 
even for solids the values of $\lambda$ are far less universal \cite{fisher1988,liu1991,majumdar1995,yeung,corberi2006,henkel,janke2007,das2014,das2015} than the growth 
exponent \cite{bray,wadhawan,onuki,binderbook1001,jones,lifshitz,furukawa1,furukawa2,binder2,
binder3,tanaka1,tanaka2,das2011,das2012,roy2012,roy2013,roy2013jcp,jung2015,nielaba,watanabe,jung2016,
roy2018,tanaka2015,janke2017,azizi,bouttes,tanaka2000} $\alpha$, the former being very strongly 
dependent upon $d$ and order parameter conservation. It is then expected that in fluids, where 
growth exponent itself is a strong function \cite{bray,furukawa1,furukawa2,binder2,binder3} of 
the space dimension and the overall composition, situation will be far more complex. 

In this work we present first results on the aging dynamics in connection with vapor-liquid phase 
separation with overall density close to the vapor branch of the coexistence curve. In addition, 
we also revisit the same problem for overall density close to the critical value \cite{das2013}. 
This is in the wake of important development \cite{das2014,das2015,das2018} in the techniques for 
analyzing aging data. For the sake of completeness, certain differences between the low and high 
density cases, with respect to structure and growth, though known, are briefly mentioned below. 

For a density close to the critical value the vapor and liquid domains percolate through the system. 
On the other hand, in the case of very low overall density the morphology consists of disconnected 
droplets of liquid phase. While hydrodynamics is important in both the cases, it manifests differently 
in the two situations, leading to different growth laws \cite{bray,furukawa1,furukawa2,binder2,binder3,das2011,das2012,roy2012,roy2013,roy2013jcp}-- overall 
coarsening rate being much faster in the higher density case. Thus, more complex picture may be 
expected with respect to the decay of autocorrelation. The objective here is to see if at least a 
power-law decay of the function is an universal feature and if so, how the exponents in the two cases 
differ from each other. For this purpose, we take help of a very effective finite-size scaling method, 
devised with the objective of analyzing simulation results for two-time quantities.

The rest of the paper is organized as follows. In section II we describe the model and provide 
details on various methods. Results are presented in section III. Finally, section IV concludes 
the paper with a summary and outlook.

\section{Model and Methods}

We consider a single component system \cite{das2011,das2012,roy2012} in which particles interact 
with each other via the Lennard-Jones (LJ) potential \cite{allen1987}
\begin {eqnarray}\label{LJ}
U(r)=4\varepsilon \left [\left (\frac{\sigma}{r} \right )^{12} - \left (\frac{\sigma}{r} \right )^6\right ].
\end{eqnarray}
Here $\sigma$ is the interparticle diameter, $\varepsilon$ is the interaction strength and $r$ is 
the scalar distance between two particles. For the sake of computational convenience we have truncated 
the potential and shifted it to zero at a distance $r=r_c$ ($=2.5\sigma$). This exercise makes the 
force discontinuous at the cut-off distance. To avoid this problem a new term is added so that the 
simulated potential \cite{allen1987} reads
\begin {eqnarray}\label{LJ2}
u(r)=U(r)-U(r_c)-(r-r_c)\frac{dU}{dr}\Big|_{r=r_c}. 
\end{eqnarray}

The phase behavior for the vapor-liquid transition in this model in various space dimensions 
have been studied \cite{das2017,errington,jia2017} via Monte Carlo simulations \cite{wild,landau}. 
In this paper we are interested in $d=3$. In this dimension the critical values of the temperature 
($T$) and the overall number density ($\rho$) are \cite{errington,jia2017} respectively 
$T_c \simeq 0.94 \varepsilon/k_B$ and $\rho_c \simeq 0.32$, where $k_B$ is the Boltzmann constant. 
For the study of aging dynamics related to the growth in droplet morphology we will consider $\rho=0.08$, 
while for the percolating case simulations will be performed with $\rho =0.3$. Of course, 
both the densities fall inside the coexistence curve at the considered values of $T$, numbers for 
which we will mention in appropriate places. 

Kinetics for the above mentioned densities have been studied via molecular dynamics (MD) simulations 
\cite{allen1987,frenkel}. We present results at multiple temperatures, values of which were 
controlled via the application of a suitable thermostat. For accurate preservation of hydrodynamics, 
a requirement for studying phase separation in fluids, one should work in microcanonical ensemble. 
However, a thermostat is needed to study kinetics for temperature driven phase separation. 
Though not perfectly, a number of thermostats are known to preserve hydrodynamics well. Examples are 
dissipative particle dynamics thermostat \cite{groot}, Nos\'{e}-Hoover thermostat (NHT) \cite{frenkel,nose}, 
Lowe-Andersen thermostat \cite{koopman}, etc. From the point of view of temperature control, 
we observe that NHT is a better one and will use it for the present study. 

All our simulations were performed in cubic boxes of side $L\sigma$. Results are presented after 
averaging over a minimum of 80 independent initial configurations for coarsening in low 
density situation and this number is 10 for the high density case.
These configurations were prepared at 
$T=10\varepsilon/k_B$, far above the critical value. Time in our MD simulations is counted in units of 
$t_0=\sqrt{m\sigma^2/\varepsilon}$, where $m$ is the mass of each particle. For the sake of convenience, 
from here on we set $\sigma$, $\varepsilon$, $k_B$, and $m$ to unity. 

We have applied periodic boundary conditions in all directions. Unless otherwise mentioned,
for $\rho=0.08$ we have simulated systems with $L=200$ and for $\rho=0.3$ we have considered $L=100$. For these
linear dimensions our systems contain 640000 and 300000 particles, respectively. Given the nature of 
the problem, that requires long runs and good statistics, these numbers are significantly large, certainly 
for MD simulations. 

From the snapshots gathered via MD simulations, average length scale, which is necessary to 
quantify $\lambda$, can be obtained by exploiting the scaling properties of various morphology 
characterizing functions \cite{bray}. These are two-point equal-time correlation function, $C(r,t)$; 
its Fourier transform, referred to as the structure factor; domain size distribution function; etc. 
E.g., the first moment of the domain size distribution function, $P(\ell_d,t)$, provides the 
average size as \cite{das2011,das2012,roy2012}
\begin {eqnarray}
\ell = \int P(\ell_d,t)~\ell_d ~ d\ell_d,
\end{eqnarray}
where $\ell_d$ is the size of a domain that can be obtained from the separation between two 
interfaces along any Cartesian direction. Equivalently, $\ell$ can also be obtained from the decay 
of $C(r,t)$, defined as $C(r,t)=\langle \psi(\vec r,t)\psi (\vec 0,t)\rangle - \langle \psi(\vec r,t)\rangle 
\langle \psi(\vec 0,t) \rangle$, via \cite{bray,das2011,das2012,roy2012, dasroy2015}
\begin {eqnarray}
C(r=\ell,t)=f,
\end{eqnarray}
where $f$ is a constant. Note here that, in order to calculate the correlation functions, $\psi$, 
at a space point $\vec{r}$, has been assigned the number $+1$ if the local density is higher than 
the critical value and we have set $\psi$ to $-1$ when this density is less than the critical number 
\cite{das2011,das2012,roy2012}. This method essentially maps the configuration to that of Ising model 
\cite{landau}. In this work we have calculated $\ell$ from Eq. (7).

\section{Results}

\begin{figure}[htb]
\centering
\includegraphics*[width=0.4\textwidth]{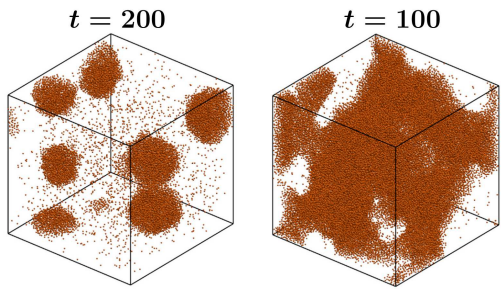}
\caption{Representative snapshots obtained during the evolutions of the single component 
Lennard-Jones systems, following quenches of high temperature homogeneous configurations to $T=0.6$. 
We have presented pictures or two different overall densities, viz., $\rho=0.08$ (left frame) and 
$0.3$ (right frame), both of which, for the chosen temperature, correspond to state points inside 
the vapor-liquid coexistence curve. The times at which the snapshots were recorded are mentioned 
in appropriate places. For both the cases we have chosen $L=64$. The quantitative results in each 
of the cases will be presented for bigger system sizes.}
\label{fig1}
\end{figure}

In Fig. \ref{fig1} we show two snapshots, one each for $\rho=0.08$ and $0.3$, obtained during 
the evolutions of the LJ systems at $T=0.6$. The morphology in the low density case consists of 
disconnected liquid drops in the vapor background \cite{roy2012}. On the other hand, for the high 
density system we have bicontinuous structure made of elongated liquid and vapor domains \cite{das2011}. 

In absence of hydrodynamics, growth in systems with conserved order parameter, which is also the 
case in the present problem, occurs with $\alpha=1/3$, referred to as the Lifshitz-Slyozov (LS) 
exponent \cite{lifshitz}. This is due to diffusive transport of material. While this mechanism is 
responsible for growth in fluids as well, dominant contribution in this case comes from hydrodynamic 
mechanisms. In fluids the growth depends upon the overall density 
\cite{furukawa1,furukawa2,binder2,binder3,tanaka1,tanaka2,das2011,das2012,roy2012,roy2013,roy2013jcp}. 
This is because the above mentioned disconnected and percolating morphologies provide different 
mechanisms, as discussed below.

In the high density case, advective transport through tube-like regions is expected to provide an 
exponent \cite{bray} $\alpha=1$, following a brief period of LS regime. This high value of $\alpha$ 
can be obtained via a balance between interfacial free energy density and viscous stress. 
Thus, this is referred to as the viscous hydrodynamic growth \cite{bray}. At a much later time, 
there is expected to be a further crossover from $\alpha=1$ to $2/3$, the latter being referred 
to as the inertial hydrodynamic regime \cite{bray}. The corresponding exponent can be obtained by 
equating the interfacial energy density with the kinetic energy density. In this paper, 
for the high density part our focus will be on the viscous hydrodynamic regime. 

Manifestation of hydrodynamics is different in the disconnected case \cite{binder2,binder3}. 
If the background vapor density is reasonably high, i.e., temperature is close to the critical value, 
the droplets may exhibit diffusive motion and the growth will primarily occur via coalescence 
among these mobile clusters. In this case, solution of a dynamical equation \cite{siggia}, 
obtained by equating the rate of change of droplet density with the collision frequency, provides 
the Binder-Stauffer (BS) \cite{binder2,binder3} growth law $\alpha=1/d$. In $d=3$, the space dimension 
of our interest, value of $\alpha$ is $1/3$, same as the LS law. However, the amplitudes in the two 
cases are expected to be different. 

\begin{figure}[htb]
\centering
\includegraphics*[width=0.4\textwidth]{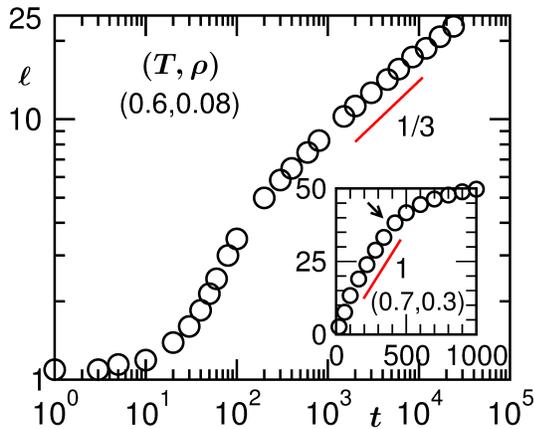}
\caption{Plots of average domain size, $\ell$, versus time, on a log-log scale, for (main frame) 
droplet ($\rho=0.08$) and (inset) percolating ($\rho=0.3$) morphologies. The data for $\rho=0.08$ 
are from $T=0.6$ and for $\rho=0.3$ the temperature is $T=0.7$. The continuous lines represent 
power-laws, exponents of which are mentioned. Here and in subsequent figures all results for 
$\rho=0.08$ and $0.3$ are presented from $L={{200}}$ and $100$, respectively, except for 
$(\rho,T)=(0.3,0.6)$. In the latter case we have used $L=96$. The arrow in the inset marks the 
onset of finite-size effects.}
\label{fig2}
\end{figure}

In Fig. \ref{fig2} we show plots of average domain size, versus time, for both types of 
morphology \cite{das2011,roy2013jcp}. For $\rho=0.08$ (main frame), we have presented data from $T=0.6$, 
whereas the temperature for $\rho=0.3$ (inset) is $0.7$. For the higher density case, 
related previous study \cite{das2013} was at $T=0.7$, which we revisit. Nevertheless, for aging 
we will also present results for $T=0.6$, for this density, at the end. From this figure, 
it can be clearly identified that growth is much faster when the overall density is higher. 
Due to this reason, for the low density case one needs to perform simulations over very long times 
to access significant scaling regime in the growth dynamics. Thus, the case of disconnected morphology 
is computationally more difficult.

On a double-log scale, the late time growth for the disconnected morphology is
consistent with $\alpha=1/3$. For the percolating case, on the other hand, an $\alpha=1$ regime 
can be identified from the plot presented in double-linear scale. The slow growth towards the end 
is related to finite-size effects. Note that for very low overall density nucleation is delayed. 
This is reflected in the reasonably long ``no-growth'' period for the disconnected pattern. 
The sharp rise of data set in this case in an intermediate time scale corresponds to the onset of 
instability and does not belong to the scaling regime of growth. Further details on the exponents 
and related aspects of growth, for this case as well as for the percolating morphology, was provided 
in earlier works \cite{das2011,das2012,roy2012,roy2013,roy2013jcp}. Next we move to our primary 
interest that is in the aging property. 

\begin{figure}[htb]
\centering
\includegraphics*[width=0.4\textwidth]{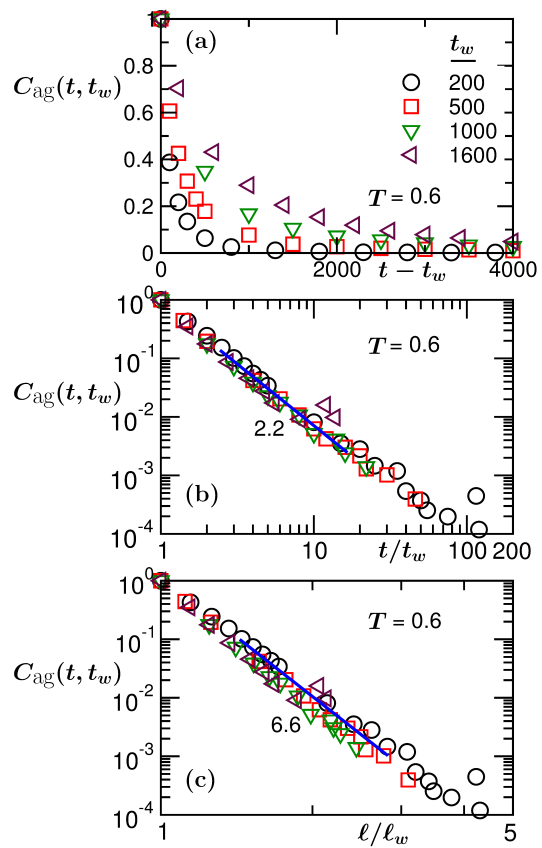}
\caption{(a) Plots of the autocorrelation function, $C_\text{ag}(t,t_w)$, versus $t-t_w$,
for the droplet morphology. Results from a few different values of $t_w$ 
are presented.
(b) Here we have plotted the correlation functions, on a log-log scale, versus $t/t_w$. 
The continuous line represents a power-law decay, exponent for which is mentioned in the 
figure. (c) Same as (b), but here the results are presented versus $\ell/\ell_w$. Again, the continuous
line is for a power-law decay. All results correspond to $T=0.6$.}
\label{fig3}
\end{figure}

In Fig. \ref{fig3}(a) we show $C_\text{ag}(t, t_w)$, with the variation of $t-t_w$, 
for $\rho=0.08$ and $T=0.6$. Results from a few values of $t_w$ are included. Clearly, unlike
the equilibrium situation, there is no overlap of data from different $t_w$, implying violation of time
translation invariance. In Fig. \ref{fig3}(b) we plot $C_\text{ag}(t,t_w)$, on a log-log scale, 
versus $t/t_w$. In this case, it appears that data 
for different $t_w$ nicely superimpose on top of each other. This collapse confirms the scaling 
behavior proposed by FH [see Eqs. (\ref{l_v_t}) and (\ref{aging2})]. The solid line in the figure 
represents a power-law decay with exponent $2.2$. The simulation data, 
over two decades in $t/t_w$, are consistent with this line. Data sets
from higher values of $t_w$ also fall on the master curve. However, given that these do not cover large
range of $t/t_w$, we abstain from presenting them.

In Fig. \ref{fig3}(c) we show again the same data sets, but here $C_\text{ag}$ is plotted as a function 
of $\ell/\ell_w$. Again, all the data sets reasonably overlap with each other. Note that we have 
chosen $t_w$ in such a way that we are practically in the scaling regime of the growth. This can be 
verified from the main frame of Fig. \ref{fig2}. Again, appearance of the data sets is linear, 
over the whole range, in the log-log scale, re-confirming a power-law decay of 
the scaling function. This master function appears consistent with the continuous line that has a 
power-law exponent $\lambda=6.6$. Noting that $\ell$ and $t$ are connected to each 
other via Eq. (1), the exponent values in Fig. \ref{fig3}(b) and 
Fig. \ref{fig3}(c) are consistent with $\alpha\simeq1/3$, confirming again the 
expected BS growth exponent in $d=3$.

Our scaled data cover a range less than a decade when plotted versus $\ell/\ell_w$.
Given that the scaling in growth appears late and the corresponding exponent is small, to reach a decade 
in $\ell/\ell_w$, without encountering finite-size effects and with acceptable statistics, it will be
necessary to run many simulations with $L\gtrsim 400$, i.e., with systems containing about five million 
particles, till about $t=10^5$. This has not been possible with the resoures available to us. 
But the presented $\ell/\ell_w$ range is comparable with the latest available studies of simplier
models \cite{das2018}, e.g., the Ising model. We emphasize again that our data cover a range 
of two decades when plotted versus $t/t_w$, in addition to the correlation function falling by
four decades.

The value $\lambda\simeq7$ certainly satisfies the FH lower bound \cite{fisher1988}, which is 1.5 
in $d=3$. But the number violates the upper bound, value of which is 3. At this point a brief 
discussion on the bounds will be useful \cite{liu1991}. A domain of size $\ell$ at time $t$ consisted 
of many domains of much smaller sizes at the waiting time  $t_w$. Since the average value of the order 
parameter within $\ell$ should be $\ell^{-d/2}$ when many domains reside within this area, one obtains 
the lower bound to be ``$d/2$''. In the special case, when no growth occurs, the upper bound 
``$d$'' is arrived at. FH themselves warned about the strictness of the upper bound. 

It is also important to mention here that Yeung, Rao and Desai (YRD) \cite{yeung} later provided a 
stricter lower bound: $\lambda \ge (d+\beta)/2$, where $\beta$ is related to the structure factor 
as \cite{wadhawan,yeung1988} $S(k \rightarrow 0, t) \sim k^\beta$, $k$ being the wave number. 
While the FH bound typically applies to growth with nonconserved \cite{bray} order parameter ($\beta=0$), 
the YRD bound is more general. This is in the sense that $\beta$ is (positive) nonzero when one deals 
with conserved \cite{bray,lifshitz} order parameter. We have checked that for the droplet morphology 
in Fig. \ref{fig1} value of $\beta$ is $\lesssim 1$ (see also Ref. \cite{paul}). This implies that the 
observed value of $\lambda$ satisfies both FH and YRD lower bounds. However, $\lambda=7$
is a number much higher than both the bounds. The huge differences can only be attributed to the dynamical aspect. 
Here we just mention that this value of $\lambda$ is in reasonable agreement \cite{das2015} with the 
corresponding number obtained for LS mechanism in $d=3$ for which also one has $\alpha=1/3$. 
Next, we present results for the percolating morphology. 

\begin{figure}[htb]
\centering
\includegraphics*[width=0.4\textwidth]{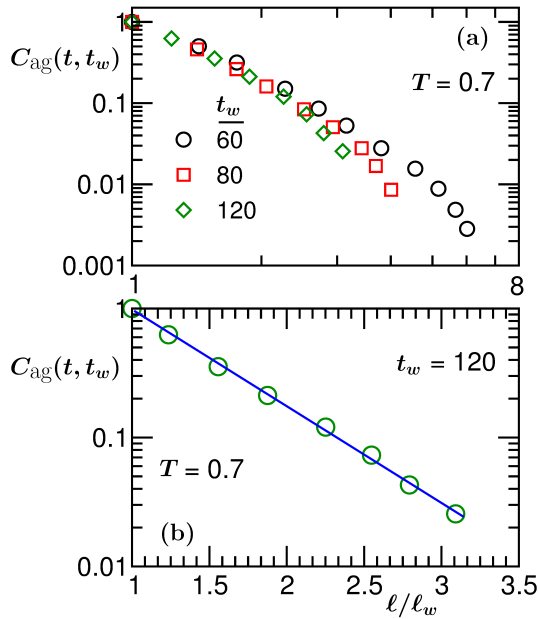}
\caption{(a) Log-log plot of $C_\text{ag}(t,t_w)$ as a function of $\ell/\ell_w$, for a few values of 
$t_w$ that are mentioned in the figure. These results are for the percolating morphology, i.e., 
from $\rho=0.3$, with $T=0.7$. 
(b) Same as (a), but in this part we have shown data for only one value of 
$t_w$, on a semi-log scale. The solid line here represents an exponential decay.}
\label{fig4}
\end{figure}

Turning to the high density case, we show log-log plots of the $C_\text{ag}(t,t_w)$ versus $\ell/\ell_w$ 
data for $\rho=0.3$ and $T=0.7$ in Fig. \ref{fig4}(a). Few different values of $t_w$, all lying in the 
linear growth regime, are used. Again, collapse of data appear nice. Deviations from the master curve 
towards the end, appearing earlier for larger $t_w$, are due to finite-size effects 
\cite{das2014,das2015,das2018}. This feature was absent in the low density case for which finite-size 
effects did not appear yet, due to much slower growth. Furthermore, in contrary to the low density case, 
the master curve, on the double-log scale, for the present density, does not appear linear. A continuous 
downward bending is clearly visible. This bending may imply \cite{ahmad2012,das2013} 
a faster than power-law decay. Thus, in Fig. \ref{fig4}(b) we show a log-linear plot \cite{das2013}, 
for a representative value of $t_w$, chosen in such a way that the chances of its falling outside the 
linear growth regime is minimal. In this plot the data indeed appear linear, and parallel to the 
continuous line that, of course, has an exponential form. However, given that the general theoretical 
prediction is that of a power-law, further analysis should be performed, if possible. A reason for the 
continuous bending could be the presence of corrections to the power-law scaling. 

Similar bending was observed in $C_\text{ag}(t, t_w)$ for coarsening during ferromagnetic ordering 
\cite{das2014,das2018}, as well as in kinetics of phase separation in solid binary mixtures \cite{das2015}. 
In both the cases it turned out that $C_\text{ag}(t,t_w)$ can be reasonably accurately written as 
($A$ and $B$ being constants) \cite{das2014,das2015,das2018}
\begin {eqnarray}\label{eq_fss}
C_\text{ag}(t,t_w)=A \text{exp}\left( \frac{-B}{x}\right)x^{-\lambda};~x=\ell/\ell_w,
\end{eqnarray}
such that one obtains a power-law only in the asymptotic limit $x =\infty$. Here also we assume the 
validity of the similar exponential correction factor and perform finite-size scaling (FSS) analysis 
\cite{das2014,das2015,das2018,landau,fisher2,barber} to check for consistency of Eq. (\ref{eq_fss}) 
with the simulation data. 

For that purpose we introduce a scaling function $Y$ that reads \cite{das2018}
\begin {eqnarray}
Y=C_\text{ag}(t,t_w) \text{exp}\left( \frac{By}{y_w}\right) y_w^\lambda,
\end{eqnarray}
where $y=L/\ell$ and $y_w=L/\ell_w$. The transformation of variables: $x \rightarrow y$ is 
motivated by the fact that an appropriate FSS variable is $L/\ell$, analogous to $L/\xi$ in 
critical phenomena \cite{landau,fisher2,barber}, $\xi$ being the equilibrium correlation length. 
Given that $y$ is a dimensionless variable, it is expected that $Y$ should be independent of system size. 
Thus, an $Y$ versus $y$ plot should have collapse of data from various different system sizes. 
It can be easily verified that in the thermodynamic limit of $y \rightarrow \infty~(L >> \ell)$, 
i.e., when there are no finite-size effects, Eq. (10) will provide the behavior in Eq. (9) if 
\begin {eqnarray}\label{eq_fss2}
Y \sim y^\lambda.
\end{eqnarray}
In the other limit, i.e., when $y \rightarrow 0 ~(t >> t_w)$, we expect $C_\text{ag}(t,t_w)$ 
to vanish, which must lead to the null value of $Y$ as well. Observation of the latter will require 
long simulations which we avoid.

An interesting fact about the FSS analysis for the aging phenomena is that one can get away 
without simulations of different system sizes \cite{das2015,das2018}. As seen in Fig. \ref{fig4} (a), 
where we have presented data for a fixed value of $L$, finite-size behavior for higher values of $t_w$ 
appear earlier in $\ell/\ell_w$, analogous to quicker emergence of the effects for smaller 
system sizes if $t_w$ were kept fixed \cite{das2015,das2018}. This is because different ``effective'' 
system sizes are available for growth when relaxation is studied by starting from different ages 
of the system. Thus, $t_w$ essentially serves the purpose of $L$ and so, we will look for collapse 
by using data from different values of $t_w$, by fixing $L$. For an optimum collapse we will adjust 
the values of $\lambda$ and $B$. 

\begin{figure}[htb]
\centering
\includegraphics*[width=0.4\textwidth]{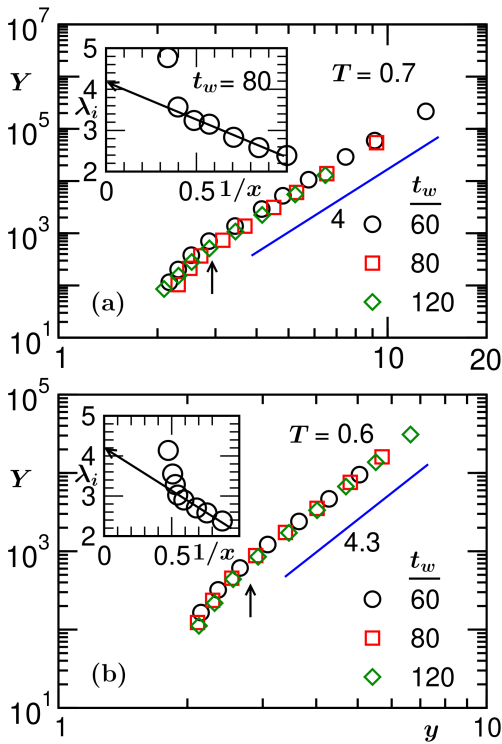}
\caption{(a) Finite-size scaling plot of the autocorrelation data for the percolating morphology. 
Here we have presented data from $T=0.7$. The solid line represents a power-law. Onset of 
finite-size effects has been marked by an arrow. Inset: Plot of instantaneous exponent $\lambda_i$ 
as a function of $\ell_w/\ell$. The solid line is a guide to the eye. (b) Same as (a) but for $T=0.6$.}  
\label{fig5}
\end{figure}

In Fig. \ref{fig5}(a) we show FSS exercise for percolating morphology data from $T=0.7$. 
The data collapse is very good and obtained for $\lambda = 4$. For large $y$, the master curve is 
quite consistent with the behavior in Eq. (11). The deviation from this power-law behavior is marked 
on the figure. This occurs when $\ell \simeq 0.37 L$. As we have previously demonstrated \cite{das2012} 
with growth data, this fraction of $L$ corresponds to the onset of finite-size effects \cite{das2012}. 
Consistency of $Y$ with the power-law all the way till the finite-size effects emerge is suggestive 
of the fact that the exponential factor indeed nicely sums up the corrections, as in a number of 
other systems, in addition to pointing to the possibility of a power-law decay of $C_\text{ag}(t,t_w)$ 
in the asymptotic limit, as opposed to an exponential decay \cite{das2013}. Results from similar 
exercise for $T=0.6$ is presented in the main frame of Fig. \ref{fig5} (b) which we describe below. 

In this case best collapse is obtained from $\lambda=4.3$. Again, for large $y$ behavior of $Y$ is 
consistent with Eq. (\ref{eq_fss2}). Appearance of finite-size effects is in agreement with that 
of higher temperature data in Fig. \ref{fig5} (a). If there exists any temperature dependence in 
$\lambda$, further studies will be needed to confirm that. Given that the values quoted in 
Figs. \ref{fig5} (a) and \ref{fig5} (b) differ from each other by a little more than $5\%$, 
we are not inclined to make a comment on this issue. Since with the variation of temperature we expect 
only change in the amplitude of growth law, it will be surprizing if the aging exponent changes due to the 
variation in amplitude only. Nevertheless, we do not rule out such a possibility, though very unlikely.

The fact on the full form can be further appreciated from the plot of instantaneous exponent \cite{das2014,huse}
\begin {eqnarray}
\lambda_i=-\frac{d~\text{ln}C_\text{ag}(t,t_w)}{d~\text{ln}x},
\end{eqnarray}
versus $1/x$, in the insets of Fig. \ref{fig5}. The linear behavior in these data sets, when 
incorporated in Eq. (12), in fact provides Eq. (9). Extrapolation of these data to $x = \infty$, 
discarding the finite-size affected parts, also lead to essentially the same values of $\lambda$ 
that we have obtained from the FSS analyses. The quoted values of $\lambda$ for this morphology satisfy 
the YRD bound $3.5$. Note here that for percolating structure in $d=3$, value \cite{yeung1988} of $\beta$ is $4$.

\section{Conclusion}

We have presented results on aging phenomena, in the context of vapor-liquid phase transitions, 
from molecular dynamics simulations of a single component Lennard-Jones fluid \cite{das2011,das2012,roy2012}. 
Effects of hydrodynamics on aging have been investigated for two different coarsening mechanisms: 
(i) growth via advective transport through tube-like liquid domains in interconnected morphology; 
(ii) equilibration through diffusive coalescence process that occurs in disconnected pattern where 
isolated liquid droplets are immersed in the matrix of the vapor phase. 

We have quantified the aging dynamics via calculation of the two-time order-parameter autocorrelation 
function \cite{wadhawan,fisher1988} $C_\text{ag} (t,t_w)$, where $t$ and $t_w$ are respectively the 
observation and waiting times. It has been demonstrated that in both the cases $C_\text{ag} (t,t_w)$ 
exhibits nice scaling with respect to $\ell/\ell_w$, the ratio of characteristic lengths at $t$ 
and $t_w$. For the disconnected domain pattern very robust power-law decay emerges from the beginning 
and the estimated value of the exponent $\lambda$ is $\simeq 7$. 

On the other hand, for the connected case there is no simple power-law, at least for $\ell/\ell_w<\infty$. 
Even though standard method of analysis indicates that the decay could be exponentially fast, an 
advanced analysis via a recently devised finite-size scaling technique \cite{das2014,das2018} is 
supportive of an asymptotic power-law decay. For small values of $\ell/\ell_w$, this method suggests, 
there exist corrections that can be consolidated in an exponential factor, similar to the observations 
in studies related to ordering in ferromagnets \cite{das2014,das2018} and phase separation in solid 
binary mixtures \cite{das2015}. 

Interestingly, the exponent $\lambda\simeq7$, for the droplet morphology, even though differs
from the interconnected counterpart of the present work, is in nice agreement with that for solid 
binary mixture \cite{das2015} in the same space dimension $d=3$. In the latter case, albeit due to a different 
mechanism, the growth exponent $\alpha$ is the same as the disconnected morphology part of this work. 
Nevertheless, one should be careful about drawing conclusion on expectation of similar values of $\lambda$ 
if $\alpha$ remains same. Value of $\alpha$ remains unaltered \cite{bray,lifshitz} for solid 
binary mixtures even in $d=2$. But we have observed a drastically different value of $\lambda$ there 
\cite{das2015}. Furthermore, the growth exponent for the percolating morphology case here is in nice 
agreement with certain active matter systems \cite{belmonte,das2017jcp}. However, aging exponents differ, 
despite being in the same space dimension \cite{das2017jcp}. Thus, the universality in the value of 
$\lambda$ is weaker than $\alpha$ and the issue is more complex. Nevertheless, it appears that at least 
the asymptotic power-law behavior is rather universal with respect to the decay of autocorrelation 
in nonequilibrium processes. Also, interestingly enough, the pre-asymptotic corrections are nicely 
summed up in an exponential factor for apparently different systems. These include coarsening in 
ferromagnets, solid binary mixtures, and now in fluids.

For both the morphologies, the Fisher-Huse (FH) lower bound is satisfied. Of course, the Yeung, Rao and 
Desai bounds appear stricter. Nevertheless, for disconnected morphology the value of $\lambda$ is 
far higher than any of the bounds. From the construction of the lower and higher bounds of FH it is 
expected that the value of $\lambda$ should be smaller for faster growth. Nevertheless, drawing general 
conclusion is difficult, given that both structure and dynamics play important roles that are till 
now unclear to us. It perhaps makes better sense to compare the percolating morphology outcome of 
this work with that of solid binary mixtures \cite{das2015} in $d=3$. This is because in both the cases 
one has similar structure, viz., value of $\beta$ is $4$. In that case we should expect a slower decay 
of the autocorrelation in fluids. This indeed is the case, making the conclusion of the work more plausible. 
Furthermore, in the above mentioned active matter problem \cite{das2017jcp} even though the growth exponent 
and space dimension are same as the percolating case of this work, $\lambda$ is smaller in the active 
matter case. This could probably be justified by the fact that $\beta$ in the latter case is smaller. 
Another general trend that we observe is that for same growth mechanism and structure, with the increase 
of space dimensionality value of $\lambda$ increases. To make general conclusions from these observations 
we need to study more cases and require appropriate theoretical considerations.

An important question one may ask here: whether there will be a jump from $\lambda\simeq7$
to $\simeq 4$, or this change will be realized in a continuous manner with the increase of density. 
Answer to this question, we believe, will be decided by whether the growth exponent $\alpha$ has a smooth
dependence on density or not. As long as the density is less than the percolation threshold, the growth
will occur via coalescence mechanism and the value of  $\alpha$ will be $1/3$, keeping $\lambda$ high.
However, with the increase of density it is possible that the onset of disconnected morphology will be delayed. As density
increases towards the percolation value, initially there may be random non-spherical structures, providing
the possibility of connectedness at early time. This feature will go away at late time, as long as
the value of density remains below the above mentioned threshold. 

The scaling method without the need for performing simulations with different system sizes is important. 
This becomes possible due to the fact that for different values of $t_w$ (or $\ell_w$) different 
effective system sizes are available for the relaxation to occur. In this sense the ratio $\ell_w/L$, 
rather than $L$, is only important. This outcome certainly helps avoiding time consuming simulations 
like molecular dynamics for fluid phase separation. 

Acknowledgement: Parts of the simulation results are obtained by using the package LAMMPS \cite{lammps}. 
AB acknowledges research fellowship from the Council of Scientific and Industrial Research, India.

$*$das@jncasr.ac.in

\end{document}